\documentclass[12pt,preprint2,natbib,iop]{emulateapj} 
\usepackage{graphicx,epsfig,amsmath,multirow,hyperref} 
\usepackage[hyphenbreaks]{breakurl}

\lefthead{Maseda et al.}
\righthead{Dynamical and Stellar Masses for
Extreme Emission Line Galaxies}
\slugcomment{Version: \today}
\begin{document}

\newcommand{\kms}{\>{\rm km}\,{\rm s}^{-1}}
\newcommand{\esca}{\>{\rm erg}\,{\rm s}^{-1}\,{\rm cm}^{-2}\,{\rm \AA}{}^{-1}}
\newcommand{\reff}{r_{\rm{eff}}}
\newcommand{\msol}{M_{\odot}}
\newcommand{\zsol}{Z_{\odot}}
\newcommand{\inverse}[1]{{#1}^{-1}}
\newcommand{\invvar}{\inverse{C}}
\newcommand{\dd}{{\rm d}}

\title{Confirmation of Small Dynamical and Stellar Masses for
Extreme Emission Line Galaxies at z$\sim$2}

\author{Michael V. Maseda\altaffilmark{1}, Arjen van der Wel\altaffilmark{1},
Elisabete da Cunha\altaffilmark{1}, Hans-Walter Rix\altaffilmark{1}, Camilla Pacifici\altaffilmark{2},  Ivelina Momcheva\altaffilmark{3}, Gabriel B. Brammer\altaffilmark{4}, Marijn Franx\altaffilmark{5}, Pieter van Dokkum\altaffilmark{3}, Eric F. Bell\altaffilmark{6}, Mattia Fumagalli\altaffilmark{5}, Norman A. Grogin\altaffilmark{4}, Dale D. Kocevski\altaffilmark{7}, Anton M. Koekemoer\altaffilmark{4}, Britt F. Lundgren\altaffilmark{8}, Danilo Marchesini\altaffilmark{9}, Erica J. Nelson\altaffilmark{3}, Shannon G. Patel\altaffilmark{5}, Rosalind E. Skelton\altaffilmark{10}, Amber N. Straughn\altaffilmark{11}, Jonathan R. Trump\altaffilmark{12},  Benjamin J. Weiner\altaffilmark{13}, Katherine E. Whitaker\altaffilmark{11}, Stijn Wuyts\altaffilmark{14}}

\affil{$^1$ Max-Planck-Institut f\"ur Astronomie, K\"onigstuhl 17, D-69117
Heidelberg, Germany; email:maseda@mpia.de}
\affil{$^2$ Yonsei University Observatory, Yonsei University, Seoul 120-749, Republic of Korea}
\affil{$^3$ Department of Astronomy, Yale University, New Haven, CT 06520, USA}
\affil{$^4$ Space Telescope Science Institute, 3700 San Martin Drive, Baltimore, MD 21218, USA}
\affil{$^5$ Leiden Observatory, Leiden University, Leiden, The Netherlands}
\affil{$^6$ Department of Astronomy, University of Michigan, 500 Church Street, Ann Arbor, MI 48109, USA}
\affil{$^7$ Department of Physics and Astronomy, University of Kentucky, Lexington, KY 40506, USA}
\affil{$^8$ Department of Astronomy, University of Wisconsin, 475 N Charter Street, Madison, WI 53706, USA}
\affil{$^9$ Physics and Astronomy Department, Tufts University, Robinson Hall, Room 257, Medford, MA 02155, USA}
\affil{$^{10}$ South African Astronomical Observatory, P.O. Box 9, Observatory 7935, South Africa}
\affil{$^{11}$ Astrophysics Science Division, Goddard Space Flight Center, Code 665, Greenbelt, MD 20771, USA}
\affil{$^{12}$ University of California Observatories/Lick Observatory and Department of Astronomy and Astrophysics, University of California, Santa Cruz, CA 95064, USA}
\affil{$^{13}$ Steward Observatory, 933 N. Cherry St., University of Arizona, Tucson, AZ 85721, USA}
\affil{$^{14}$ Max-Planck-Institut f\"ur extraterrestrische Physik, Giessenbachstrasse 1, D-85748 Garching, Germany}

\begin{abstract}
Spectroscopic observations from the \textit{Large Binocular Telescope} and the
\textit{Very Large Telescope} reveal kinematically narrow lines ($\sim 50~\kms$)
for a sample of 14 Extreme Emission Line Galaxies (EELGs) at redshifts $1.4
< z < 2.3$. These measurements imply that the total dynamical masses of these
systems are low ($\lesssim 3\times10^9~\msol$). Their large [O III] $\lambda5007$ equivalent widths
($500-1100$ \AA) and faint blue continuum emission imply young ages of $10-100$
Myr and stellar masses of $10^8-10^9~\msol$, confirming the presence of a violent starburst. The dynamical masses represent the first such determinations for low-mass galaxies at $z > 1$.  The stellar mass formed in this vigorous starburst phase represents a large fraction of the total (dynamical) mass, without a significantly massive underlying population of older stars. The occurrence of
such intense events in shallow potentials strongly suggests that
supernova-driven winds must be of critical importance in the
subsequent evolution of these systems.

\end{abstract}

\section{INTRODUCTION}
The $z>1$ universe contains a remarkably large number of galaxies with
extremely luminous nebular emission lines in comparison to their faint
blue continua \citep{vdw}. These extreme emission line galaxies (EELGs) can
have [O III] and/or H$\alpha$ equivalent widths (EWs) in excess of 500 \AA$ $ \citep{atek,vdw,shim,gb2}.
Such observations suggest that young starbursts dominate the energy
output of these otherwise faint galaxies, potentially serving as the principle mode of mass build-up in low-mass galaxies.  While similar objects do exist
at $z<1$ \citep{greenpea,izotovSDSS}, they have a much lower comoving number density thereby implying that
their abundance is a strong function of time.

Without further information, the dwarf interpretation of these galaxies is merely plausible.  More massive populations of older stars could easily be outshone by
the young starbursts: an old stellar population can have mass-to-light ratios up to 50 times larger than those of the bursts in the near-IR, so the main uncertainty in the interpretation
of the observations hinges on the determination of the total masses of
these systems. Additionally, the presence of strong emission lines can hinder attempts to determine the stellar mass content, as standard SED-fitting codes do not contain emission line contributions.  Hence we do not yet understand the role of this mode of star formation in the broader context of
galaxy formation.  When these bursts occur in truly low-mass galaxies ($\sim10^8~\msol$),
the EELGs may represent the main formation mode of present-day dwarf
galaxies, as argued by \citet{vdw}.  Alternatively, if these bursts are embedded in more massive systems ($\gtrsim10^9~\msol$), we may be witnessing the early formation stage of Milky Way-type galaxies.

Accurate mass estimates are key in addressing this issue, particularly dynamical masses. For this
purpose we now present near-infrared spectroscopy of 14 EELGs at
redshifts $1.4 < z < 2.3$ with [O III] $\lambda5007$ equivalent widths $>$ 500 \AA$ $ from the
\textit{Large Binocular Telescope} (LBT) and the \textit{Very Large Telescope} (VLT). 
These are the first dynamical mass measurements of such low-mass, high-redshift
galaxies, and we also derive accurate stellar mass estimates
through stringent modeling of the continuum and emission line
measurements from CANDELS
multi-wavelength photometry \citep{candels1,candels2} and low-resolution grism spectroscopy from the
3D-HST survey \citep{gb}.

We adopt a flat
$\Lambda$CDM cosmology with $\Omega_m=0.3$ and H$_0=70 ~$km s$^{-1}$
Mpc$^{-1}$ throughout.

\section{Candidate Selection and Observations}
We select a sample of 17 objects with restframe equivalent widths $>$ 500 \AA$ $ in [O III] $\lambda5007$: five are from the photometrically-selected sample of \cite{vdw} in the GOODS-S and UDS fields, and the 12 remaining objects were selected based on their 3D-HST grism spectra in the COSMOS, GOODS-S, and UDS fields.  One object, \textit{COSMOS-10320}, although fulfilling the criteria, exhibits broad and asymmetric [O III] (and also H$\alpha$) of 240$\pm$10 $\kms$.  As this object is an obvious outlier (with a potential AGN contribution), we exclude it from the subsequent analysis and focus on the remaining 16 objects.  Although the targets are very faint in the continuum ($m_{F140,AB} \gtrsim 24$), the 
emission lines are strong, 
with fluxes $ > 10^{-17} ~\esca $, making emission line detections possible with $\sim$1 hour integrations on 8m class telescopes.   We observe five objects using long-slit observations with the X-SHOOTER wide-band spectrograph \citep{vernet} at the VLT from August to December 2012 (one slit contained two objects), focusing here on the combined $YJHK$ NIR region (1024$-$2480 nm with resolution $R \sim 5000$), although it simultaneously observes in the UV-Blue and the Visible regions.  Four had 40 minute integrations, while one object was observed for a total of 120 minutes in the near-IR over the course 
of two nights.  The 
remaining objects in the sample were observed using the LUCI1 multi-object spectrograph \citep{seifert} at the LBT with four separate masks between April 2012 and March 2013 in the $J$-, $ H$-, and/or $K$-band (depending on the 
redshift, as we targeted [O III] and/or H$\alpha$) with resolution $R = 6000 - 8000$ for a minimum of 45 minutes per band.  Two objects in the total LUCI1 sample had \textit{a priori} equivalent widths greater than 500 \AA$ $, but severe contamination from OH sky lines at the predicted position of the lines prevents a line extraction and they are not included in this sample.  In total, five objects were detected in both H$\alpha$ and [O III], one was detected only in H$\alpha$, and eight were detected only in [O III].  The faintest detected line in the X-SHOOTER (LUCI1) sample is 7.4 (6.0) $\times 10^{-17} \esca$ with signal-to-noise of 42 (2).  For all observations, seeing was better than 1$''$ and typically between 0.3$''$ and 0.8$''$.  All exposures were dithered by 3$''$ to decrease dependence on the pixel-to-pixel detector variations and defects.

\begin{figure}
\plotone{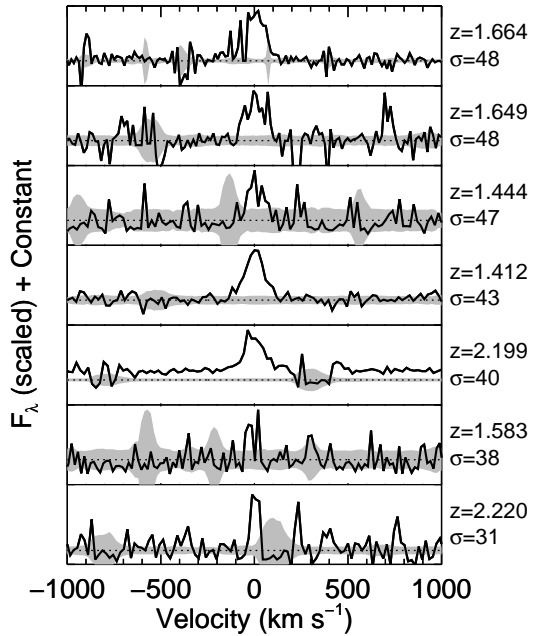}
\caption{Plot of the [O III] $\lambda$5007 emission line for each object, scaled to the peak flux value.  Gray regions show the +/- 1-$\sigma$ flux uncertainties.  Typical uncertainties are smaller than 10$^{-4}$ in redshift and $\sim$8 $\kms$ in $\sigma$.}
\label{fig:vel}
\end{figure}
\begin{figure}
\plotone{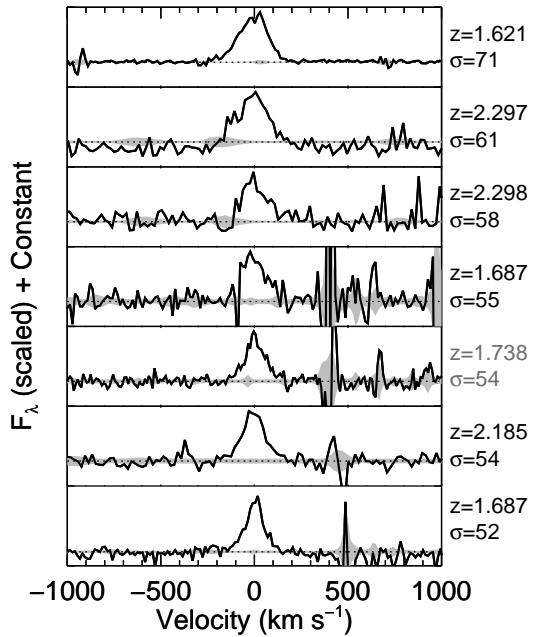}
\caption{Same as Figure \ref{fig:vel} for the remainder of the sample.  The single object with gray labels denotes H$\alpha$.}
\label{fig:vel2}
\end{figure}

Reduction of the X-SHOOTER data is performed using version 2.0.0 of the ESO XSHOOTER pipeline\footnotemark[15]\footnotetext[15]{\url{http://www.eso.org/sci/software/pipelines/xshooter/xsh-pipe-recipes.html}}, which provides merged, 2D near-IR spectra.  Reduction of the LUCI1 data is performed using a custom pipeline, with the wavelength calibration done using the OH sky lines and based on the \texttt{XIDL} routines\footnotemark[16]\footnotetext[16]{\url{http://www.ucolick.org/\textasciitilde xavier/IDL/}}.  For the brightest emission lines, we also use \texttt{XIDL} for the final sky subtraction, which uses a spline-fitting algorithm to measure and remove the sky lines.  

Identified emission lines in the 1D spectra are fit with Gaussian functions, where all lines in a subregion of the spectrum (i.e. [O III] $\lambda\lambda$4959,5007 and H$\beta$) are forced to have the same width and only the ratio of the two [O III] components is fixed to 2.98 \citep{oiii}.  When both [O III] and H$\alpha$ are observed for a single object, we take the width of the higher-S/N line complex to be the ``true'' width, which is [O III] for this entire sample.  The two line widths are always consistent within 1-$\sigma$.  A full description of the data reduction is given in Maseda et al. (in prep).

Extracted emission lines are shown in Figures \ref{fig:vel} and \ref{fig:vel2}.  The sample has a median line width of 48 km s$^{-1}$ with an average uncertainty of 8 km s$^{-1}$, after correcting for seeing and instrumental broadening which is typically $\lesssim$ 20\% of the intrinsic line width.

\section{Dynamical and Stellar Masses}

\subsection{Dynamical Mass Measurements}
\label{sec:sizes}

The velocity dispersions derived above can be used to estimate the dynamical masses according to:
\begin{equation}
 M_{dyn} = C\frac{r_{\rm{eff}}\sigma^2}{G}.
\label{eqn:dyn}
\end{equation}

Here, we have adopted the half-light radius $\reff$ as the virial
radius. We take $\reff$ as the half-light radius from \citet{vdw12}, who provide size
measurements from the F125W and F160W HST/WFC3 CANDELS imaging. We
choose the filter that does not contain the [O III] emission line to
ensure that the size is measured from the continuum light as much as
possible. In cases where H$\alpha$ is in F160W and [O III] is in F125W, we use the F160W size as [O III] is brighter and therefore may affect the broadband flux more.  For objects in which the only line is [O III] in F125W, \citet{vdw} note that the sizes measured in both bands are still consistent.  The typical $\reff$ is 1 kpc, which is larger than the HWHM of the PSF, so these sources are indeed resolved. 
As noted in \citet{weiner}, kinematic estimates using line widths yields a variety of results: \citet{rix} calculate $C=2.8$ for inclined rotating disks, while \citet{barton} calculate $C=2.1$ for blue compact dwarfs; \citet{erb06} use a simple geometric correction to obtain $C=3.4$.  Here we adopt $C=3$, with a conservative uncertainty of 33\%, as in \citet{rix}.  Note that this value of $C$ would be the same if we assume that these
systems are spherical.  We find that the 14 EELGs have $\log(M_{dyn}/\msol)$
ranging from 8.7 to 9.7, with a median of 9.1 and an average uncertainty of 0.3.

There are several potential systematic effects that may
affect these estimates. First, for these systems the measured half-light radius is not necessarily equal to the virial radius.  Indeed, some have irregular morphologies that are not well fit by single-component profiles. Second, these systems likely have an irregular dynamical structure and may not be virialized.  

\subsection{Stellar Mass Measurements}

\begin{deluxetable*}{lcccccccc}
\tabletypesize{\scriptsize}
\tablecaption{Summary of Near-IR Observations and Masses\label{tab:obs}}
\tablewidth{0pt}
\tablehead{ \colhead{ID} & \colhead{RA} & \colhead{Dec} & \colhead{Instrument} &\colhead{$z_{spec}$}&\colhead{EW$_{[O III],5007}$}&\colhead{$\sigma_{[O III]}$}&\colhead{$M_{dyn}$}&\colhead{$M_* (\texttt{MAGPHYS}$)} \\
    & (deg) & (deg) & &&(\AA) &($\kms$) &($\msol$) &($\msol$)}
\startdata
COSMOS-15144&150.156769 &2.360800&LUCI1 &1.412&1130$\pm$247&43.3$\pm$8.9&9.11$\pm$0.34&8.10$_{-0.26}^{+0.20}$\\
COSMOS-13848&150.176987&2.345390&LUCI1&1.444&888$\pm$351&46.7$\pm$14.4&9.22$\pm$0.40&8.58$_{-0.22}^{+0.14}$\\
COSMOS-12807&150.159546&2.333301&LUCI1&1.583&628$\pm$152&38.2$\pm$10.0&8.88$\pm$0.37&7.95$_{-0.24}^{+0.18}$\\
UDS-7444&34.473888&-5.234233&X-SHOOTER&1.621&713$\pm$42&71.1$\pm$5.7&9.66$\pm$0.33&8.78$_{-0.16}^{+0.07}$\\
COSMOS-16207&150.183090 &2.372948&LUCI1&1.649&536$\pm$20&47.7$\pm$9.5&9.40$\pm$0.34&8.43$_{-0.12}^{+0.17}$\\
UDS-3760&34.428570&-5.255318&X-SHOOTER&1.664&731$\pm$86&48.2$\pm$5.9&9.04$\pm$0.31&7.98$_{-0.09}^{+0.11}$\\
UDS-3646&34.426483&-5.255770&X-SHOOTER&1.687&701$\pm$95&54.7$\pm$6.1&9.47$\pm$0.33&8.51$_{-0.13}^{+0.12}$\\
GOODS-S-17892&53.171936 &-27.759146&X-SHOOTER&1.687&693$\pm$47&52.3$\pm$5.7&9.05$\pm$0.30&8.95$_{-0.11}^{+0.10}$\\
GOODS-S-26816&53.071293&-27.705803&X-SHOOTER&1.738&861$\pm$66&54.4$\pm$4.5\tablenotemark{a}&8.86$\pm$0.31&8.53$_{-0.11}^{+0.09}$\\
UDS-11484&34.431400 &-5.212120&LUCI1&2.185&723$\pm$95&54.2$\pm$9.4&9.35$\pm$0.34&8.97$_{-0~~~}^{+0~~~}$\\
COSMOS-11212&150.124237 &2.313672&LUCI1&2.199&598$\pm$189&40.3$\pm$8.9&8.78$\pm$0.36&8.77$_{-0.26}^{+0.23}$\\
COSMOS-8991&150.095352 &2.287247&LUCI1&2.220&714$\pm$85&30.9$\pm$9.0&8.65$\pm$0.40&9.05$_{-0.27}^{+0.21}$\\
UDS-14655&34.391373 &-5.195310&LUCI1&2.297&503$\pm$34&61.0$\pm$10.8&9.67$\pm$0.33&9.37$_{-0.31}^{+0.11}$\\
UDS-4501&34.390755 &-5.250803 &LUCI1&2.298&803$\pm$162&57.8$\pm$9.7&9.07$\pm$0.33&8.32$_{-0.19}^{+0}$\enddata
\tablecomments{All IDs refer to the CANDELS catalog for that particular field (COSMOS, UDS, or GOODS-S), all equivalent widths are quoted in the restframe, and all masses are log quantities.}
\tablenotetext{a}{H$\alpha$ width.}
\end{deluxetable*}

\label{sec:sed}
With confirmed redshifts, measured EWs of multiple lines, and
multi-wavelength photometry, we are now in a position to estimate the
stellar masses and improve upon the photometry-only method of \citet{vdw}. We take 0.3$-$2.2$\mu$m photometry for the two objects in the
GOODS-S field from \citet{goodss} and the six objects in the UDS
field from \citet{uds}.  Visual inspection of the IRAC Ch. 1/2 images reveal that eight out of 14 objects have bright neighboring objects that contaminate the flux measurements.  For consistency we perform our analysis without IRAC flux measurements for any of the objects, but we note that for those with uncontaminated IRAC fluxes, our modeling results (see below) do not change significantly.  That is, the available IRAC fluxes do not reveal an underlying, older population of stars.  No such multi-wavelength photometry is
as of yet available for the six objects in the COSMOS field. For these
objects we use CANDELS 4-band HST photometry (ACS F606W and F814W, WFC3 F125W and F160W).


Here we fit the broadband spectral energy distributions, including line fluxes measured from 3D-HST grism spectroscopy, of our galaxies using a custom version of the \texttt{MAGPHYS} code\footnotemark[17]\footnotetext[17]{\url{http://www.iap.fr/magphys/magphys/MAGPHYS.html}} \citep{magphys} that includes nebular emission computed using the Pacifici et al. (2012) model (C. Pacifici et al., in prep.).  The stellar emission is computed using the latest version of the \citet{bc} models using a \citet{chabrier03} IMF, and the attenuation by dust is accounted for using the two-component prescription of \citet{cf}. The nebular emission is computed using the \texttt{CLOUDY} photoionization code \citep{cloudy}, as described in \citet{charlotlonghetti,pacifici12}. \texttt{MAGPHYS} uses a Bayesian approach to compare the measured photometry of observed galaxies with an extensive library of 100,000 spectral 
energy distribution models spanning a wide range in star formation histories, ages, and metallicities.  The standard \texttt{MAGPHYS} priors (calibrated using more massive galaxies at low redshift) are not optimized for this specific population of young ages and low metallicities, so we have modified the standard priors to include a larger fraction of low metallicities (between 0.025 and 1 Z$_{\odot}$), and younger ages by allowing both rising and declining star formation histories, all with superimposed random bursts of star formation.  This method results in stellar masses in the range $\log(M/\msol) = 8.0-9.4$, which are listed in Table \ref{tab:obs}.

\begin{figure}
\plotone{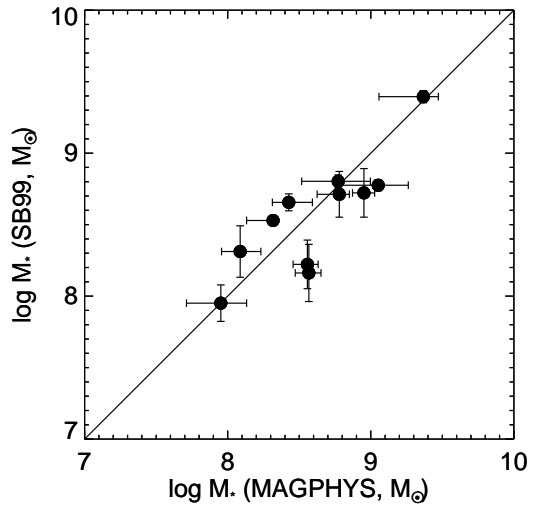}
\caption{Comparison of \texttt{MAGPHYS}- and \texttt{Starburst99}-derived stellar masses for our sample.  \texttt{Starburst99} utilizes the equivalent width of H$\beta$ (determined from photometry alone) to calculate the masses, while \texttt{MAGPHYS} utilizes the full photometric SED and the emission line fluxes.}
\label{fig:sb99}
\end{figure}

\citet{vdw} estimated stellar masses based on photometry alone, making simplistic assumptions for the star formation history, emission line properties, and the metallicity.  In Figure \ref{fig:sb99} we compare our stellar mass estimates with those estimated using the photometric method.  Our values are 1.1 times larger (median) with a scatter of 0.20 dex, consistent with no systematic offset.  The \texttt{MAGPHYS} modeling results reinforce the notion that these galaxies
are dominated, in terms of stellar mass, by a very young stellar
population.    While the \texttt{MAGPHYS} modeling
 uses much more information, the crucial elements in both mass estimates are the blue
 continuum and the strong emission lines, which strongly constrain any
 modeling approach.  


\begin{figure}
\plotone{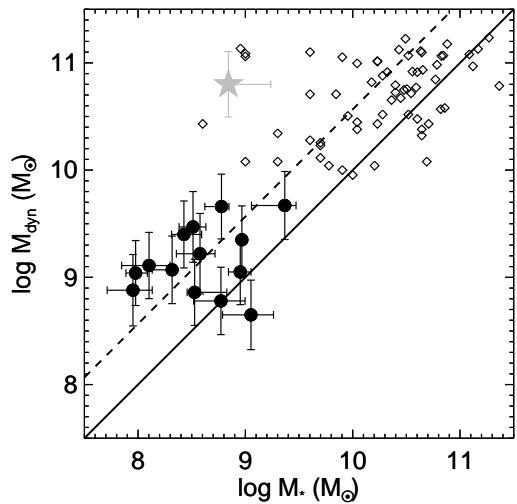}
\caption{Dynamical masses determined from the velocity width of the emission lines versus stellar masses from the \texttt{MAGPHYS} SED fits to the full optical/near-IR SEDs for our equivalent width-selected sample.  The dashed line shows the average value of 27.1\% of the total dynamical mass made up by stars.  The gray point is \textit{COSMOS-10320}, which is not considered in the analysis.  Open diamonds are from \citet{erb06} for star-forming galaxies at $z \sim 2$.  Although the $M_{dyn}$ values were derived in different manners (see Sec \ref{sec:sizes}), the relationship between $M_{dyn}$ and $M_{\star}$ is similar for the two samples.}
\label{fig:mstarmdyn}
\end{figure}

Figure \ref{fig:mstarmdyn} compares the \texttt{MAGPHYS} stellar mass estimates with the
dynamical estimates. $\log(M_{dyn}/M_{\star})=$ 0.57 (27\% of the total mass is in stars) $\pm$ 0.21 (random) $\pm$ 0.34 (systematic)
for the sample where the 0.34 dex systematic uncertainty is from the
dynamical mass (see Section 2). The 0.21 dex random uncertainty
contains the contributions from the measurement uncertainties and the
limited sample size.  The three points closest to the $M_{dyn} = M_{\star}$ line illustrate the challenges to any modeling approach.
Two of them are the only $z \sim 2.2$ galaxies from the COSMOS sample, where the 4-band CANDELS photometry does not sample any continuum redward of [O III] (one of which is also severely contaminated by an OH sky line, making our line dispersion estimate more of a lower limit), and the third is an object with two distinct components in the WFC3 imaging, where the assumptions contained in the dynamical mass estimate may not accurately reflect the true conditions in the system.

The low dynamical masses confirm the low-mass nature of these systems directly and exclude the presence of large amounts of
unseen stars, gas, dust, or dark matter that exceed the observed
amount of stellar matter by more than a factor of five.  Our implied maximal gas fractions do not exceed those for more massive galaxies
at similar redshifts, which range from $\sim 30 - 80\%$ \citep{daddi10,tacconi}.  As seen in Figure \ref{fig:mstarmdyn}, our galaxies have similar $M_{dyn}/M_{\star}$ ratios to the starforming sample of \citet{erb06}, albeit with EWs (and hence specific star formation rates) that are a factor of four higher.  

\section{Concluding Remarks}
In this Letter, we show kinematic line widths in the range $30-70~\kms$
for a sample of 14 EELGs (with EW $> 500$ \AA) at redshifts $1.4 < z <
2.3$. This constitutes the first direct mass measurements for such galaxies at these epochs, with total masses $\sim
10^{9.1}~\msol$.  SED modeling results in stellar masses $\sim 10^{8.5}~\msol$, ruling-out the presence of an evolved, massive stellar population.  Therefore, we conclude that these nascent galaxies are undergoing intense starbursts, and the stars produced in the single burst contribute substantially to their total mass budget. This confirms that the abundant population of EELGs at $z>1$ 
demonstrate a common starburst phase among low-mass galaxies at these epochs, the intensity of which has only recently been reproduced by hydrodynamical simulations \citet{shensims}.  While the contribution of such strong starbursts to the growth in stellar mass over cosmic time depends on their duty cycle, which is so far unconstrained observationally, their ubiquitous nature at these redshifts \citet{vdw} points towards the brief starburst phase as important in the mass build-up of most (if not all) dwarf galaxies.  

Given the intensity of the starbursts and the shallow potential wells
in which they occur, supernova-driven winds likely dominate the star
formation history and subsequent evolution of these systems \citep{larson}.
The starbursts may affect the central dark
matter distribution \cite[e.g,][]{n96,rg,pontzen,zolotov}
and produce cored profiles that are commonly observed in present-day, low-mass galaxies.
For a review see \citet{deblok}, and \citet{walker11,amorisco} for recent advances.
Our current data set does not allow us to make stronger conclusions about the presence of feedback and winds via asymmetric or separate broad/narrow components in individual galaxies.  However, with future spectroscopic studies of these objects, we will be able to search for such signals in stacked spectra.

In the present-day universe, such extreme starbursts are very rare
\cite[e.g.][]{greenpea}, but at early epochs ($z>4-6$) such events may
well be the rule rather than the exception. It is becoming
increasingly clear that strong emission lines affect the search for
and interpretation of high-z galaxies. Strong emission line galaxies
at moderate redshifts ($z\sim2$) can masquerade as drop-out selected $z>10$
candidates \cite[see discussion in e.g.,][]{coe,bouwens,ellis,gb13}.
Furthermore, for true high-redshift galaxies these strong emission lines are likely omnipresent \citep{smit} and affect
the broad-band SED, so they should therefore be included in
the modeling as described here in Section \ref{sec:sed} \cite[also see][]{curtislake,schaerer}. However, the results presented here are encouraging.
We suggest that if strong emission lines are evident, then it is
likely that the total stellar mass does not greatly exceed the mass of the
young stellar population traced by the blue continuum.

\acknowledgements
MVM is a member of the International Max Planck Research School for Astronomy
and Cosmic Physics at the University of Heidelberg, IMPRS-HD, Germany.  This work is based on observations taken by the 3D-HST Treasury Program and the CANDELS Multi-Cycle Treasury Program with the NASA/ESA HST, which is operated by the Association of Universities for Research in Astronomy, Inc., under NASA contract NAS5-26555, and at the European Southern Observatory, Chile, Program 089.B-0236(A).

{\it Facilities:} \facility{LBT}, \facility{VLT:Melipal}, \facility{HST}.
\bibliographystyle{apj}

\clearpage

\clearpage

\end{document}